\documentclass[twoside,slac_one]{revtex4}
\usepackage{graphicx}
\usepackage{fancyhdr}
\usepackage{amsmath} % American Mathematics Society standards
\usepackage{bm}% bold math
\usepackage{amsxtra}
\usepackage{amssymb}
\usepackage{amsthm}
\usepackage{latexsym}
\usepackage{lscape}
\usepackage{subfigure}
\usepackage{xspace}

\newcommand{\ppbar}{$p\bar{p}$}
\newcommand{\fb}{fb$^{-1}$}
\newcommand{\gevcc}{GeV/c$^2$}
\newcommand{\Hgg}{$H\rightarrow\gamma\gamma$}
\newcommand{\BrHgg}{B(\Hgg)}
\newcommand{\bbar}{$b\bar{b}$}
\newcommand{\Mgg}{\mbox{$M_{\gamma \gamma}$}\xspace}

\begin{document}

\title{Search for the Higgs boson in \Hgg\ decays in \ppbar\ collisions \linebreak[1] at 1.96 TeV }

\author{K. R. Bland}
\affiliation{Department of Physics, Baylor University, Waco, Texas, USA}
\author{(on behalf of the CDF and D0 Collaborations)}

\begin{abstract}
Recent searches conducted at the Fermilab Tevatron for the Higgs boson in the diphoton 
decay channel are reported using 7.0~\fb\ and 8.2~\fb\ of data collected at the CDF and D0 
experiments, respectively. Although the standard model (SM) branching fraction is small, the 
diphoton final state is appealing due to better diphoton mass resolution compared with dijet 
final states. In addition, other models --- such as fermiophobic models where the Higgs does 
not couple to fermions --- predict much larger branching fractions for the diphoton decay. 
Here, results are presented for both a SM and fermiophobic Higgs boson as well as a SM 
search based on a combination of the CDF and D0 analyses.
\end{abstract}

\maketitle

\thispagestyle{fancy}

%%%%%%%%%%%%%%%%%%%%%%%%%%%%%%%%%%
\section{Introduction}

The standard model (SM) of particle physics has proven to be a robust
theoretical model that very accurately describes the properties of 
elementary particles observed in nature and the forces of interaction
between them. In this model, the electromagnetic and weak forces
are unified into a single electroweak theory. The measured
masses of the particles that mediate the electroweak force, however, are 
vastly different --- the photon has zero mass while the $W$ and $Z$ 
bosons have masses almost 100 times heavier than the mass of a proton. 
To explain this difference, the theory predicts the existence
of a Higgs field which interacts with the elecroweak field via electroweak
symmetry breaking to produce masses for the $W$ and $Z$ bosons
while leaving the photon massless. 
Interaction with the Higgs field would also explain
how other fundamental particles acquire mass.
An additional spin-0 particle, the Higgs boson, is also 
predicted to arise from the Higgs field.
This particle is the only SM particle that has not been observed in nature
and evidence of this boson would be a direct test of the theory. 

The Higgs mechanism is predicted to give mass to other particles, yet the mass of the Higgs boson itself 
is a free parameter of the theory that must be determined experimentally. Direct searches at the Large
Electron-Positron Collider (LEP) at CERN and indirect electroweak measurements result in a preferred
SM Higgs boson mass $M_H$ between 114.4 and 185~\gevcc\ at 95\% confidence level (C.L). In this 
region, the range 156~$<M_H<$~177~\gevcc\ has additionally been excluded at 95\% C.L. by
direct searches at the Fermilab Tevatron \ppbar\ Collider~\cite{Tev:2011cb} and the range
above $M_H>$~146 (145)~\gevcc\ has been excluded at 95\% C.L. by direct searches at the ATLAS (CMS) 
experiment from the $pp$ Large Hadron Collider (LHC) at CERN~\cite{CMS_Aug2011,ATLAS_Aug2011}.

\begin{figure}[ht!]
\makebox[\textwidth]{
\subfigure[~Gluon fusion]{\includegraphics[scale=0.3]{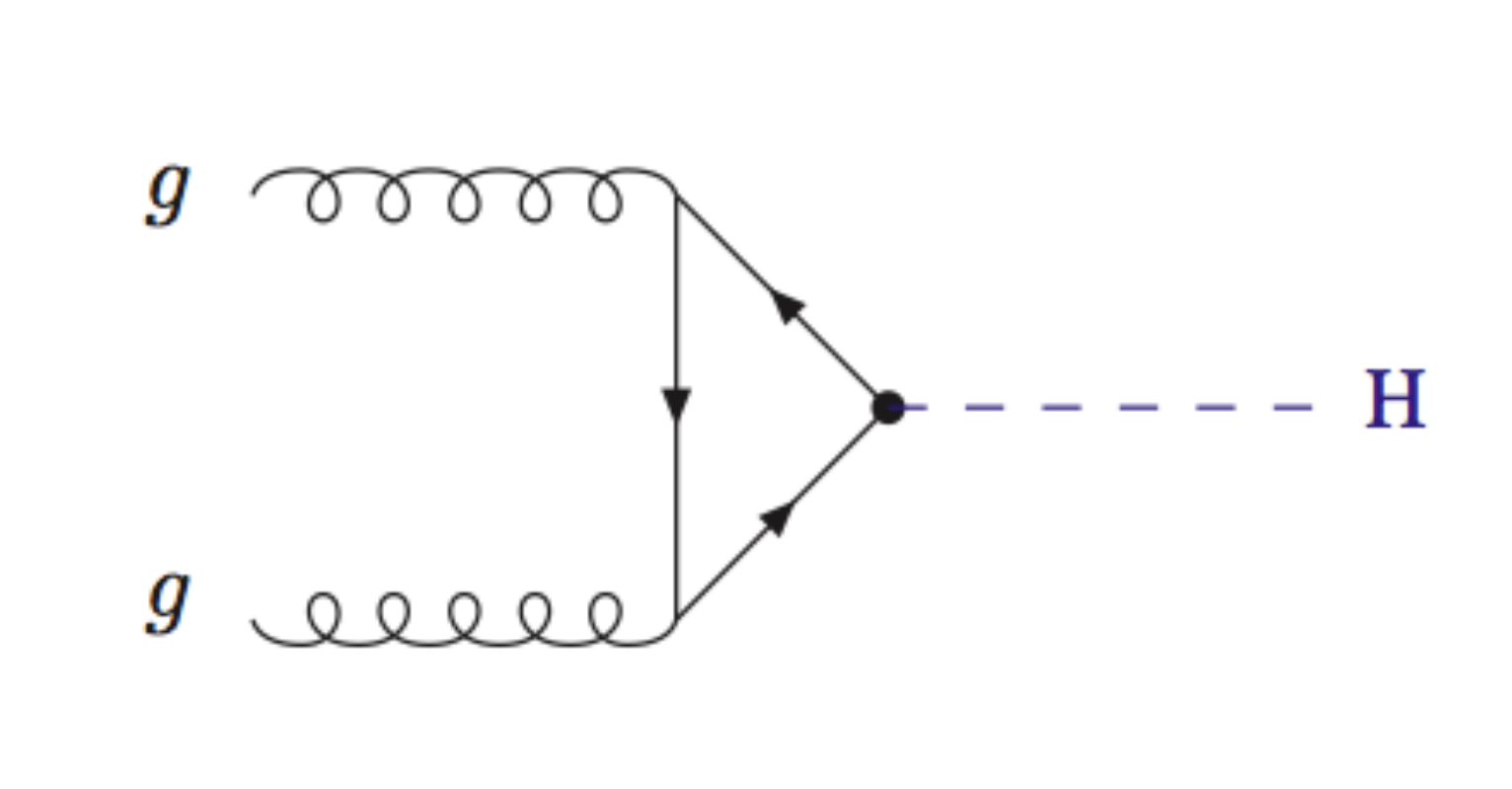}}
\subfigure[~Associated production]{\includegraphics[scale=0.3]{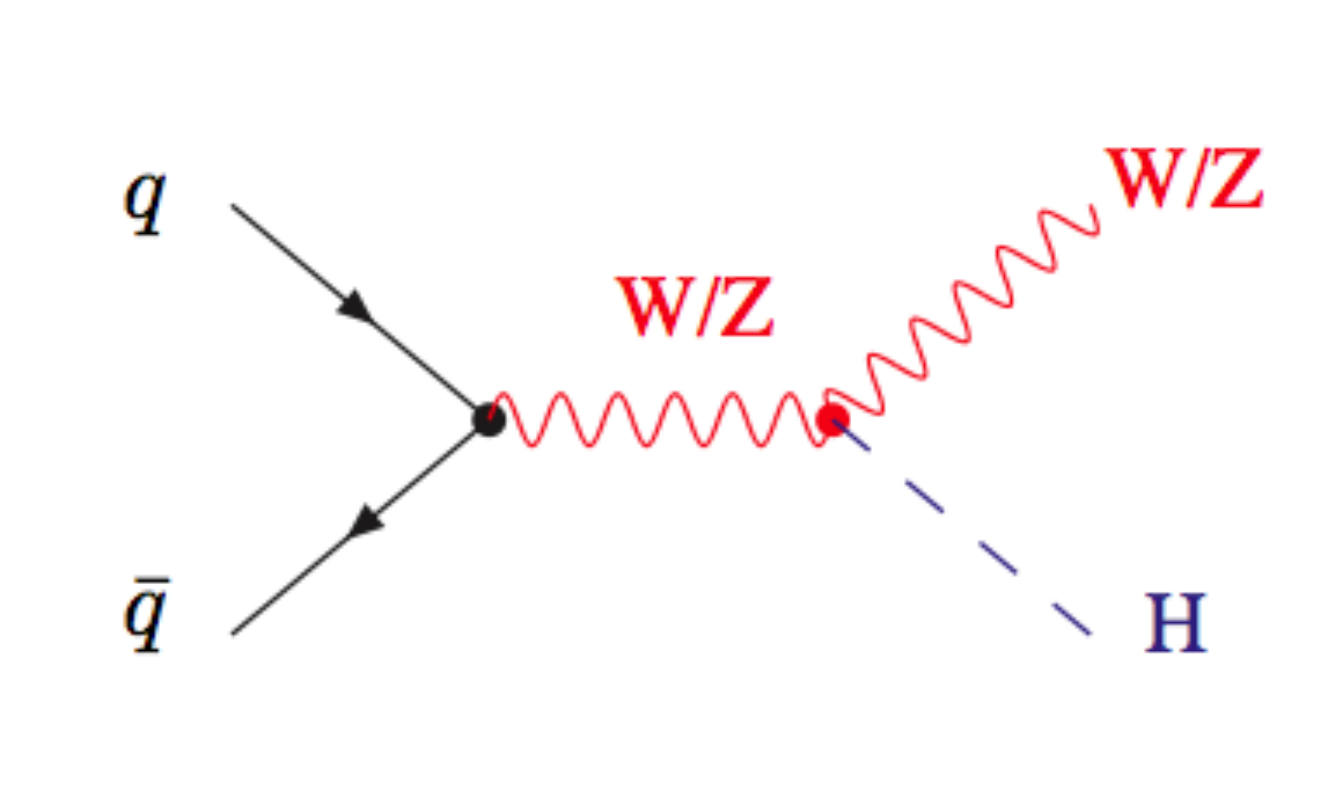}}
\subfigure[~Vector boson fusion]{\includegraphics[scale=0.3]{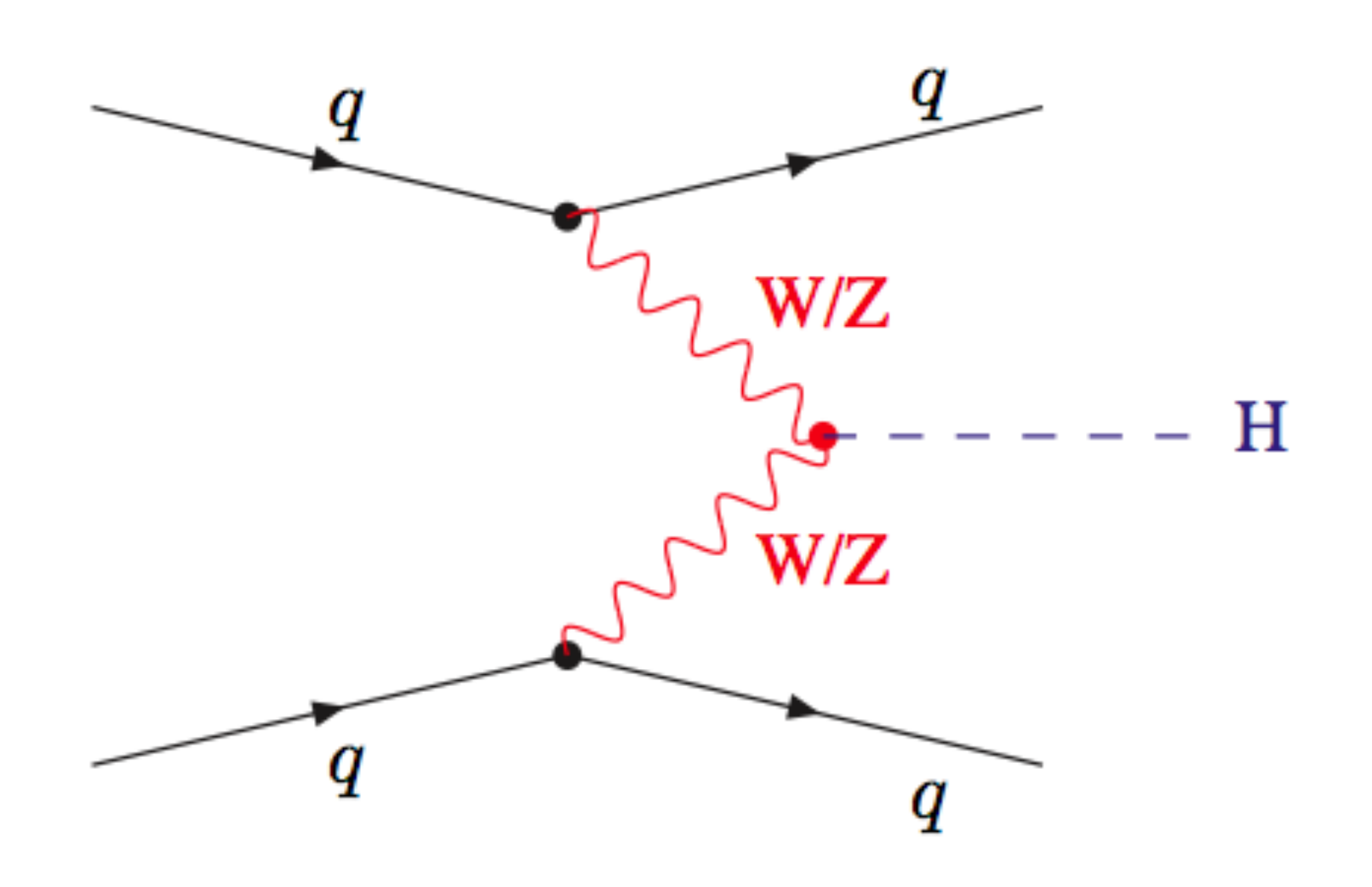}}
}
\caption{The dominant production mechanisms at the Tevatron for the SM Higgs boson. 
For the fermiophobic benchmark model, SM couplings are assumed, however the gluon fusion 
process is suppressed and is therefore not included.} \label{production}
\end{figure}

At the Tevatron, the Higgs boson would be produced most often
through gluon fusion ($gg \rightarrow H$), followed by 
associated production with 
either a $W$ or $Z$ vector boson 
($q\bar{q}\rightarrow VH$, $V=W,Z$) and vector boson fusion ($q\bar{q}\rightarrow q^{\prime}\bar{q}^{\prime}H$). Figure~\ref{production} shows diagrams 
of these processes and 
Table~\ref{tab:sigma} gives the predicted SM cross
sections for Higgs boson masses between 100 and 150~\gevcc.
The SM Higgs boson search strategy at the Tevatron is driven by the most 
dominant decay modes. At lower mass ($M_H<$~135~\gevcc), $H \rightarrow$~\bbar\ provides
the greatest sensitivity to Higgs boson observation despite the exclusion of the gluon
fusion process for this channel due to large multijet backgrounds.
For $M_H>$~135~\gevcc, $H \rightarrow WW$ provides the greatest sensitivity.
Further sensitivity to a SM Higgs observation at the Tevatron is obtained by the 
inclusion of more challenging channels such as $H \rightarrow \gamma \gamma$.  

\begin{table*}[b]
\caption{SM Higgs boson production cross sections for several $M_H$ values~\cite{D0Higgs2011} 
are shown for gluon fusion, associated 
production and vector boson fusion (see also Fig.~\ref{production}). The fermiophobic Higgs 
boson benchmark model assumes SM couplings, however the gluon fusion process is 
suppressed and is not included. 
The branching fractions for the decay to a photon pair are shown for both the SM Higgs boson
and benchmark fermiophobic Higgs boson, calculated from {\sc hdecay}~\cite{Djouadi:1997yw}.}\label{tab:sigma}
\begin{center}
\makebox[\textwidth]{
\begin{tabular}{c|c|c|c|c|c|c}
\hline
$M_{H}$ (GeV/c$^2$)                        &  
$\sigma_{gg \rightarrow H}$ (fb)           &
$\sigma_{WH}$ (fb)           &
$\sigma_{ZH}$ (fb)           & 
$\sigma_{VBF}$ (fb)                     & 
$B$($H \rightarrow \gamma \gamma)$ (\%) &
$B$($H_f \rightarrow \gamma \gamma)$ (\%) \\
\hline
100	&	1821.8	&	291.9	&	169.8	&	100.1	&	0.15	&	18.5	\\
110	&	1385.0	&	212.0	&	125.7	&	85.1	&	0.19	&	6.03	\\
120	&	1072.3	&	150.1	&	90.2	&	72.7	&	0.21	&	2.33	\\
130	&	842.9	&	112.0	&	68.5	&	62.1	&	0.22	&	1.07	\\
140	&	670.6	&	84.6	&	52.7	&	53.2	&	0.19	&	0.54	\\
150	&	539.1	&	64.4	&	40.8	&	45.8	&	0.14	&	0.27	\\ \hline                    
 \end{tabular}}
\end{center}
\end{table*}

The SM branching fraction for a Higgs boson decaying to a photon pair \BrHgg\ is very small, reaching
a maximal value of about 0.2\% at $M_H$~=~120~\gevcc\ (see Table~\ref{tab:sigma}). 
The greatest sensitivity for this channel, however, is obtained for 110~$<M_H<$~140~\gevcc, in the 
preferred region from electroweak constraints and in a region where combined Tevatron Higgs boson
searches are least sensitive~\cite{Tev:2011cb}.
The diphoton final state is also appealing due to its cleaner
signature compared to $b$ jets. The better reconstruction efficiency for photons provides a larger 
relative acceptance of \Hgg\ events and the photon's better energy resolution leads to a narrow 
\Mgg mass peak for the Higgs boson, which is a powerful discriminant against smoothly falling 
diphoton backgrounds. 
These experimental signatures help make the diphoton final state one of the most promising search modes 
for Higgs boson masses below 140~\gevcc\ at ATLAS and CMS experiments at the LHC,
which have recently presented first results in this channel~\cite{CMSPho_Aug2011,ATLASPho_Aug2011}.

In addition to SM \Hgg\ production, one can devise many possible
Beyond the Standard Model (BSM) scenarios where \BrHgg\ is
 enhanced.\footnote{An informative summary of the various models that modify \BrHgg\ can
be found in Reference~\cite{Mrenna}.}  
Any resonance observed could also 
then be evidence for a BSM Higgs. 
In the SM, the spontaneous symmetry breaking mechanism requires
a single doublet of a complex scalar field. However, it is likely that 
nature does not follow this minimal version and that a multi-Higgs sector
may be required. Here, we also consider a model which requires a doublet Higgs
field for which the symmetry breaking mechanism responsible for giving
Higgs masses to gauge bosons is separate from that which generates
the fermion masses. In the benchmark model considered, a ``fermiophobic'' 
Higgs boson ($H_f$) is predicted that assumes SM couplings to bosons and 
vanishing couplings to all fermions. The gluon fusion process is then suppressed
and only VH and VBF processes remain, which results in a reduction in the
production cross section by a factor of four. This reduction is 
compensated, however, by the branching fraction for this model, which can be larger
than that predicted by the SM scenario by more than two orders of magnitude for low Higgs
boson masses (see Table~\ref{tab:sigma}). The higher branching fraction causes a larger number of 
predicted fermiophobic Higgs boson events compared to the SM Higgs boson.
Direct searches at LEP set a lower limit on the fermiophobic Higgs boson mass
of 109.7~\gevcc\ with 95\% C.L. 

Here, we present a search for both a SM
and fermiophobic Higgs boson in the diphoton final state
from \ppbar\ collisions at $\sqrt{s} =$~1.96~TeV from the Fermilab Tevatron Collider. 
An inclusive sample of diphoton data are collected by the D0 and CDF
experiments, corresponding to an integrated luminosity of 8.2 and 7.0~\fb,
respectively. By combining the results from each analysis, Tevatron limits 
on the SM cross section multiplied by \BrHgg\ are also presented relative
to the SM prediction.

%%%%%%%%%%%%%%%%%%%%%%%%%%%%%%%%%%
\section{Prompt Photon Signature}

The dominant backgrounds to prompt photons originating from the event vertex
are electrons faking photons and jets faking photons. 
The latter is more frequent and typically occurs when a
jet fragments into a $\pi^0$ or $\eta$ meson which then decays
to multiple photons. These delayed photons are collinear and are often
mis-reconstructed as a single photon.  
In order to identify high-energy prompt photons and reduce
these backgrounds, both the D0
and CDF analyses start by searching for photon candidates with the 
following signature:
(i) the majority of the energy should be deposited in the EM
calorimeter rather than the hadronic calorimeter, (ii) the EM cluster should
be isolated in the calorimeter, (iii) there should be no high-momentum tracks 
originating from the primary event vertex that are associated with the EM cluster, 
and (iv) the EM shower profile is consistent
with that of a prompt photon. More details on this identification
can be obtained from the primary references, Ref.~\cite{D0Higgs2011} 
for D0 and Ref.~\cite{CDF_SM2011,CDF_BH2011} for 
CDF.\footnote{See Ref.~\cite{CDF_2011_Hgg} for CDF results with 
updated systematic uncertainties on the expected signal.}

\section{D0 Analysis}

\begin{figure}[ht!]
\makebox[\textwidth]{
\includegraphics[scale=0.3]{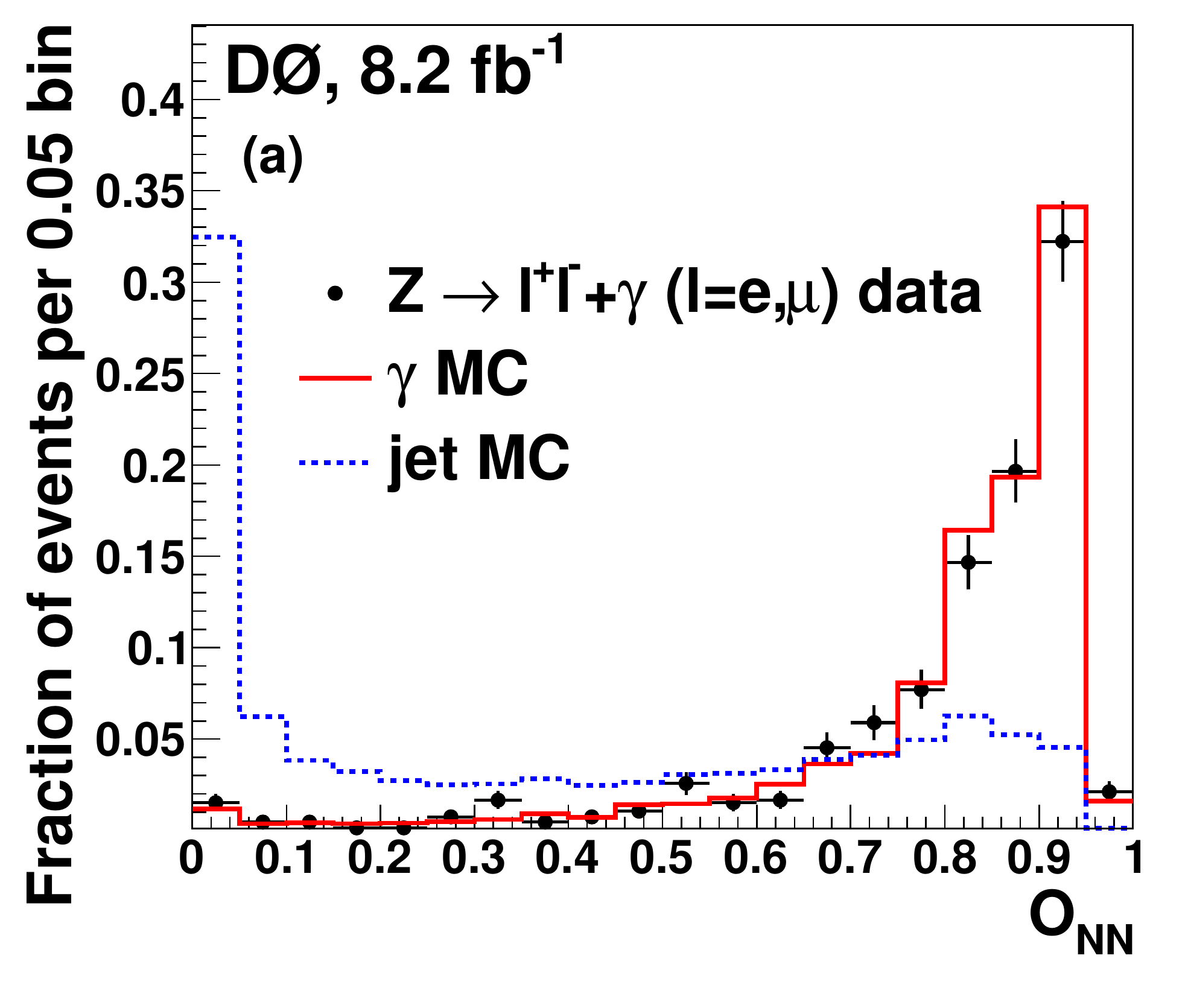}
\subfigure{\includegraphics[scale=0.3]{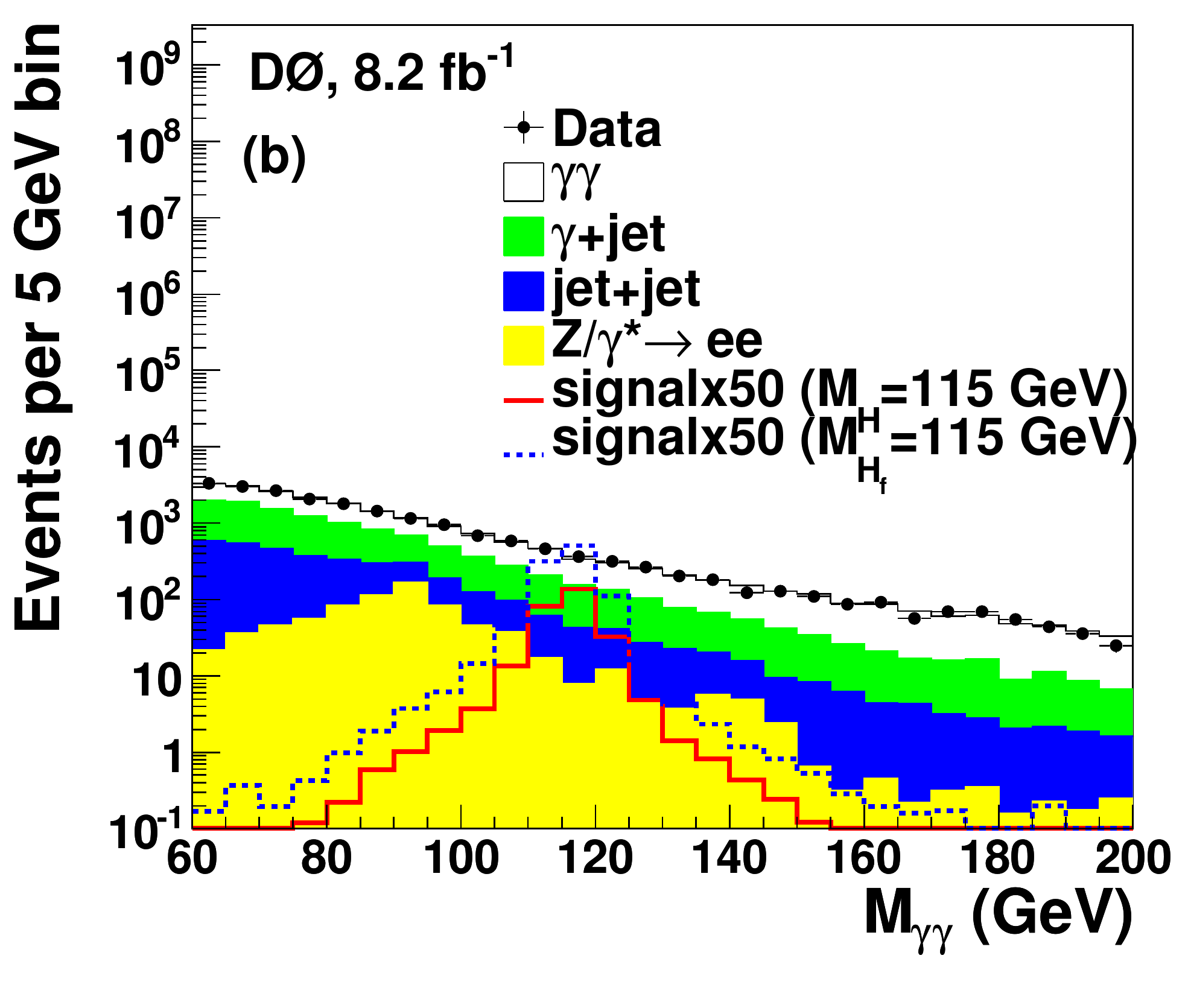}}
\subfigure{\includegraphics[scale=0.3]{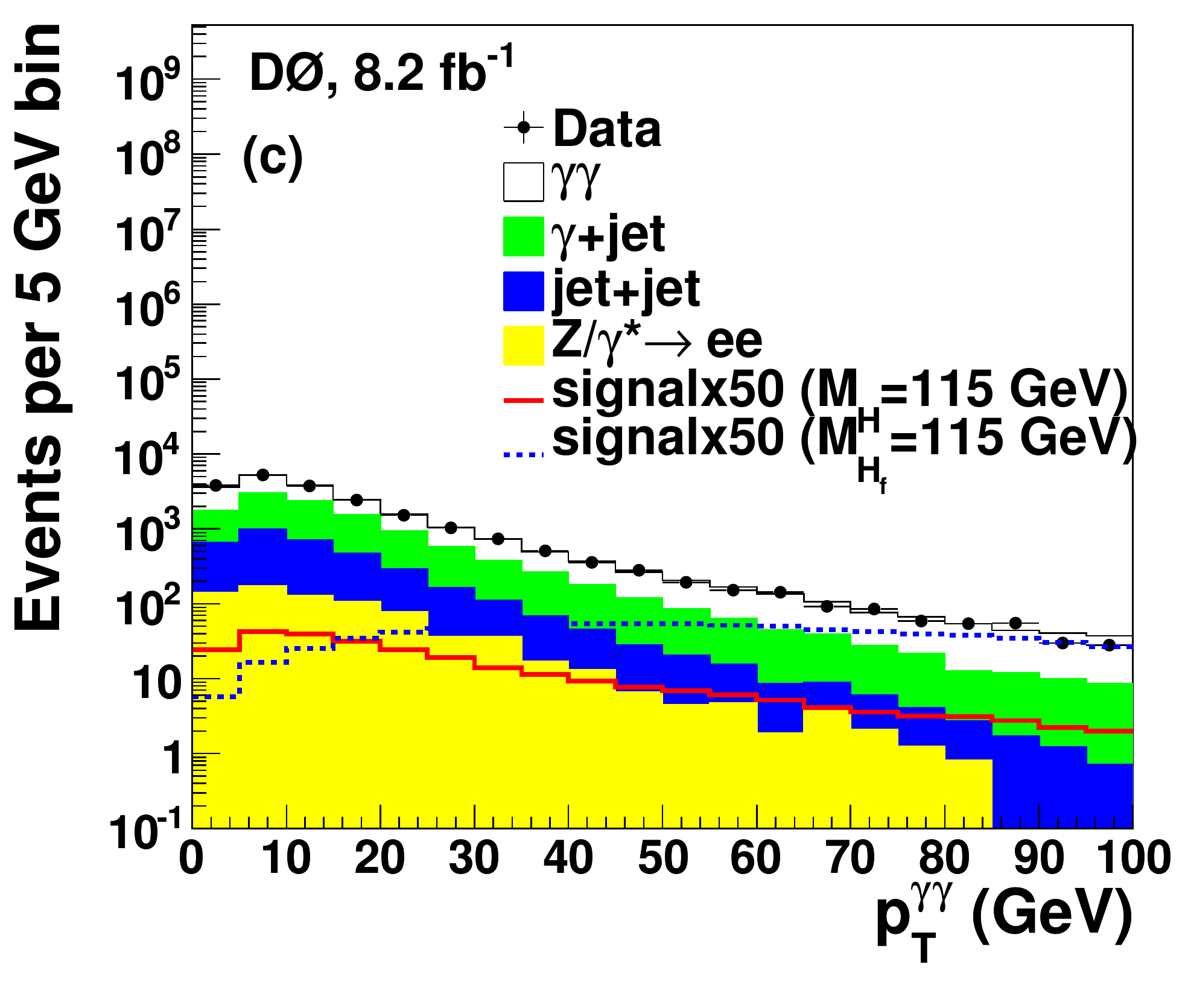}}
}
\caption{
(a) Neural net output response for photon candidates from diphoton MC, jet MC, 
and radiative $Z$ boson decays in the data. The signal photons have an $O_{NN}$
peaked towards 1 and jet backgrounds have an $O_{NN}$ peaked towards 0.
For diphoton events that pass the full selection, 
the (b) diphoton mass and (c) diphoton $p_T$ with signal shapes
(for a Higgs boson mass of 115~\gevcc)
are shown for both the SM and fermiophobic scenario. These are two of the five
kinematic variables used as inputs for the BDT training (see Fig.~\ref{D0_kin_BDT}).} \label{D0_kin_input}
\end{figure}

At D0, we select events with at least two photon candidates 
with $|\eta|<$~1.1 and transverse
momentum $p_T>$~25~GeV/c. In addition to the basic photon selection
outlined above, a Neural Network (NN) is used to further discriminate
jet backgrounds from prompt photons. This NN discriminant is trained
using photon and jet Monte Carlo (MC) samples and constructed
from well-understood detector variables sensitive to differences
between photons and jets. 
The output for this discriminant (O$_{NN}$) 
is shown in Fig.~\ref{D0_kin_input} for the photon and jet MC samples,
where the signal peaks near one and the background near zero. 
The photon efficiency $\epsilon_{\gamma}$ for any cut on this output is determined
from a true photon sample in the data obtained from 
$Z \rightarrow l^+l^-\gamma$ events ($l = e$ or $\mu$).
The jet efficiency $\epsilon_\textrm{jet}$ (fraction of jets
misidentified as a photon) is determined
from a sample of jets misidentified as photons in the data.
A cut of 0.1 is applied which retains more than 98\% of true photons
and rejects 40\% of misidentified jets. 

The two highest $p_T$ photons that pass this selection are used to form the diphoton
system and the diphoton mass $M_{\gamma\gamma}$ is then required to be greater
than 60 GeV/c$^2$. The difference in azimuthal angle of the two photons $\Delta \phi^{\gamma\gamma}$ 
is also required to be greater than 0.5, which
keeps 99\% of the Higgs boson signal but reduces 
prompt QCD photons originating 
from fragmentation, a process not well modeled in the simulation. 

For a Higgs boson signal decaying to two photons, data with the above selection
is composed of both an irreducible and a reducible background. The irreducible
background is from two SM QCD photons from the hard interaction where
the shape for different kinematic variables is modeled from {\sc sherpa} MC 
and the normalization of this background is obtained from a fit made
to the final discriminant distribution when setting limits on Higgs
boson production. The reducible background
is composed of fake events where at least one photon candidate is 
misidentified as a prompt photon. 
Both the shape and normalization of kinematic distributions for Drell-Yan 
$Z/\gamma^*\rightarrow e^+e^-$ events are obtained from 
{\sc pythia} MC prediction. The shape for $\gamma+$jet and jet+jet
background distributions is obtained from an independent data sample
selected from diphoton events that pass
all other photon selection requirements but have a reversed requirement
on the $O_{NN}$ output for one or both photon
candidates. The normalization for this sample is determined
from the data using a 4$\times$4 matrix method. 
For each data event that passes the full selection, a 4-component vector is constructed 
$(w_{pp},w_{pf},w_{fp},w_{ff})$ where the value of one element is 1 and other 
elements are 0 based on whether one or both photon candidates pass
a stronger requirement of $O_{NN}>0.75$. The weight 
$w_{pp}$ ($w_{ff}$) then represents events
where both photon candidates pass (fail) and 
$w_{pf}$ ($w_{fp}$) represents events where
only the leading (subleading)\footnote{The leading
photon refers to the highest $p_T$ photon
candidate in the event and subleading refers to the second highest.}
photon candidate passes.
The efficiency of this cut for the
photon and jet samples ($\epsilon_{\gamma}$ and $\epsilon_\textrm{jet}$)
are parametrized as a function of $\eta$ and used to construct
a 4$\times$4 efficiency matrix $\mathcal{E}$.
A system of linear equations
\begin{equation}
(w_{pp},w_{pf},w_{fp},w_{ff})^T=\mathcal{E}\times(w_{\gamma\gamma}, 
w_{\gamma j},w_{j\gamma},w_{jj})^T
\label{lineq}
\end{equation}
is then solved on an event-by-event basis
in order to obtain a weight for events with two true photons 
$w_{\gamma\gamma}$, a weight for events with two jets
faking a photon $w_{jj}$, and a weight for events
where only the leading (subleading) candidate 
is a true photon $w_{\gamma j}$ ($w_{j\gamma}$).
Then, for example, the number of dijet events is taken as the 
sum of the dijet weights obtained for each data event:
\begin{equation}
N_{jj}=\displaystyle\sum\limits_{i=1}^{N_{data}} w^i_{jj}.
\label{sum}
\end{equation}
Similarly, the $\gamma +$jet background is determined
from the sum of the $w_{\gamma j}$ and $w_{j\gamma}$
components. 

The resulting background composition is then determined 
for each Higgs boson mass hypothesis ($M_H\pm30$~GeV)
where the approximate percentage of $\gamma\gamma$ events is 53\%, 
of $\gamma j+jj$ events is 44\%, and of Drell-Yan events is 3\%. 
The simulated shape of the Higgs boson signal is generated
separately for each production process and normalized using
cross section and branching fraction values shown in 
Table~\ref{tab:sigma}. See the primary D0 reference for more
detail on signal expectation (Ref.~\cite{D0Higgs2011}).

\begin{figure}[ht!]
\makebox[\textwidth]{
\subfigure{\includegraphics[scale=0.3]{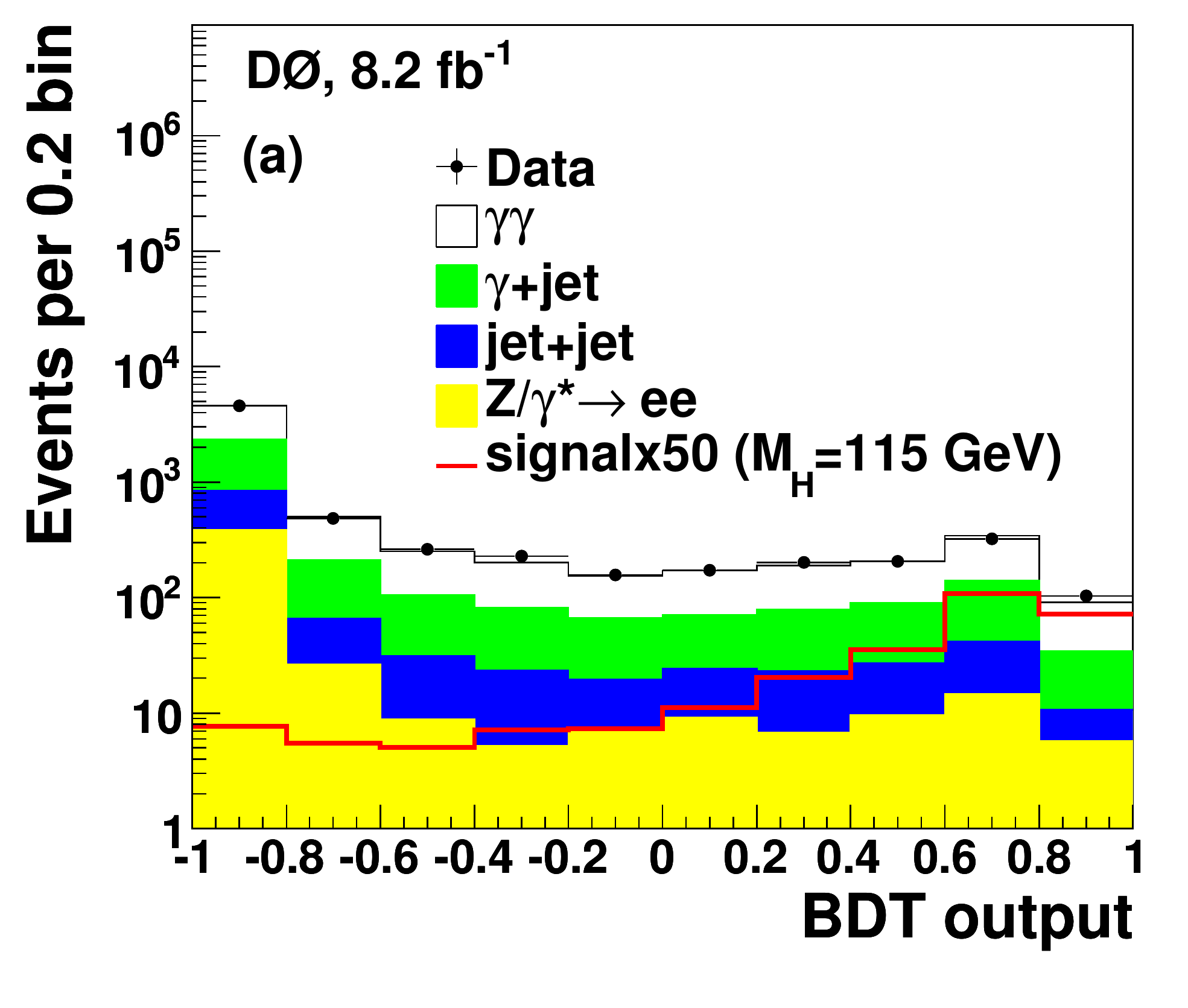}}
\subfigure{\includegraphics[scale=0.3]{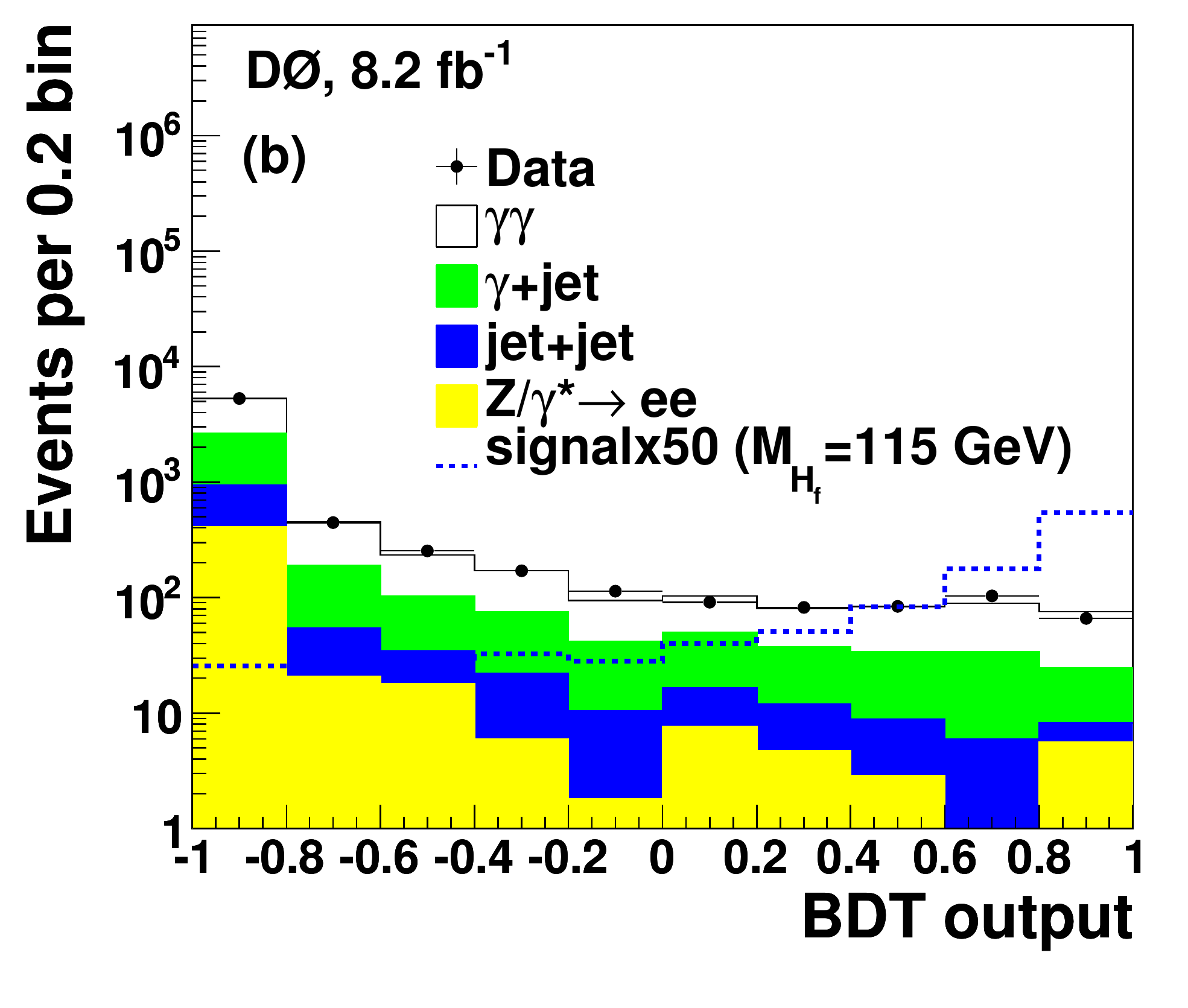}}
}
\caption{The BDT output distributions for the (a) SM and (b) fermiophobic
Higgs boson searches for a Higgs boson mass of 115~\gevcc. The data
and background predictions are shown, along with the signal prediction
multiplied by a factor of 50.} \label{D0_kin_BDT}
\end{figure}

The most recent searches for a Higgs boson in the diphoton 
final state exploit the narrow $M_H$ resolution
to discriminate between the signal and background. 
With a good understanding of the background modeling
for different kinematic distributions, however,
further sensitivity is gained at D0 by using these 
variables with a multivariate technique.
In addition to the diphoton mass $M_{\gamma\gamma}$,
four other variables with kinematic differences between the
signal and background
are also considered for this analysis:
the transverse momentum of the diphoton system $p_T^{\gamma\gamma}$, 
$\Delta\phi_{\gamma\gamma}$, and the transverse momentum of the leading 
and subleading photon, $p_T^1$ and $p_T^2$ respectively. These five well-modeled
kinematic variables
are used to construct a single discriminant from a boosted decision
tree (BDT) algorithm, trained to distinguish a Higgs boson
signal from the backgrounds. This process is repeated
for each Higgs mass hypothesis for 100~$<M_H<$~150~\gevcc\ in 2.5~\gevcc\ steps, 
and is trained separately for a SM and fermiophobic Higgs boson signal.
As an example, the $M_{\gamma\gamma}$ and $p_T^{\gamma\gamma}$ distributions
($M_H=$~115~\gevcc) for the data, background, and expected SM and
fermiophobic Higgs boson signal are shown in Fig.~\ref{D0_kin_input}, along with 
the resulting BDT response for both the SM and fermiophobic scenario 
(Figure~\ref{D0_kin_BDT}). 

%%%%%%%%%%%%%%%%%%%%%%%%%%%%%%%%%%
\section{CDF Analysis}

At CDF, the leading two photons  are required 
to have $p_T>15$~GeV/c. Plug photons ($1.2<|\eta|<2.8$) are
selected using a standard ID used at CDF~\cite{Aaltonen:2009in}.
Central photons ($|\eta|<1.05$) are identified in a similar manner as in the D0 analysis,
using a NN output constructed from variables sensitive to distinguishing
prompt photons from jet backgrounds. This NN discriminant, used
for the first time at CDF in this analysis, 
increases photon signal efficiency by 5\% and background rejection
by 12\% relative to standard photon ID at CDF. 
Photon efficiencies for both central and plug photons are validated 
using Z~$\rightarrow e^+e^-$ decays in 
both data and MC. 

$H\rightarrow\gamma\gamma$ signal acceptance is further increased by
reconstructing events in which a single central photon converts into an electron-positron pair, 
which is found to occur approximately 15\% of the time for $|\eta|<1.05$.
A base set of selection requirements 
is applied that searches for a central electron\footnote{``Electron'' is used to refer to either 
$e^+$ or $e^-$.} with a colinear, oppositely signed track nearby.\footnote{The EM object with a larger (smaller) $E_T$ track associated with it is referred to as the primary (secondary) electron.} The proximity of the two electron tracks is determined from their $r-\phi$ separation at the radius of the conversion and from the difference in $\cot \theta$ of the two tracks, where $\cot \theta = p_z/p_T$. The tracks from both electrons are required to point to a fiducial electromagnetic energy cluster. Photons of a higher $p_T$ range are selected by requiring the secondary electron to have $p_T>$~1.0~GeV/c and the reconstructed conversion photon to have $p_T>$~15~GeV/c. In order to reject jet backgrounds, only a small fraction of hadronic $E_T$ associated with the primary electron's cluster is allowed. Additionally, requirements are made on the conversion candidate's calorimeter isolation which is obtained from the primary electron's isolation energy~\cite{Aaltonen:2009in} with the secondary electron's $p_T$ subtracted if its track points to a different calorimeter phi tower. The shape describing the ratio of transverse energy to transverse momentum ($E/p$) is peaked at one for isolated photon conversions, but has a long tail for photon conversions from $\pi^0$ or $\eta$ $\rightarrow \gamma\gamma$ decays due to the extra energy from the unconverted photon. Restrictions on this ratio then provide a further way to remove jet backgrounds. The conversion $E_T$ is obtained from the primary electron's $E_T$ with the secondary electron's $p_T$ added if it is in a different calorimeter tower while the photon's reconstructed transverse momentum is obtained by adding the vector sum of the two track's momenta at the radius of the conversion. A final requirement removes events with a small radius of conversion, primarily to reduce prompt electron-positron pairs from Dalitz decays of neutral pions $\pi^0~\rightarrow~e^+e^-\gamma$.
The direction of the conversion photon's momentum is obtained by taking the vector
sum of the individual track momenta; however, better \Hgg\ mass resolution is obtained
by setting the total momentum to be the conversion's energy obtained from EM calorimeters,
which additionally constrains the photon's mass to zero.
Reconstruction of photon conversions in the CDF analysis provides an improvement of 
about 13\% in sensitivity to a Higgs boson signal.

Data events in the CDF analysis are divided into four 
independent subsamples according the position and type 
of the photon candidate. In CC events (the most sensitive
category), there are two photons
in the central region of the detector. In CP events,
one photon is in the central region and one is the plug region.
 For each of these categories, the two highest $p_T$
photons in the sample are selected. If a CC or CP event is not identified,
then two additional categories are considered.
In C$^{\prime}$C events, both photons
are central but one has converted and is reconstructed from
its $e+e-$ decay products. Finally, in C$^{\prime}$P events, 
one photon is in the plug region and the other is a 
central conversion photon. 

\begin{figure}[bh!]
\begin{center}
\includegraphics[scale=0.4]{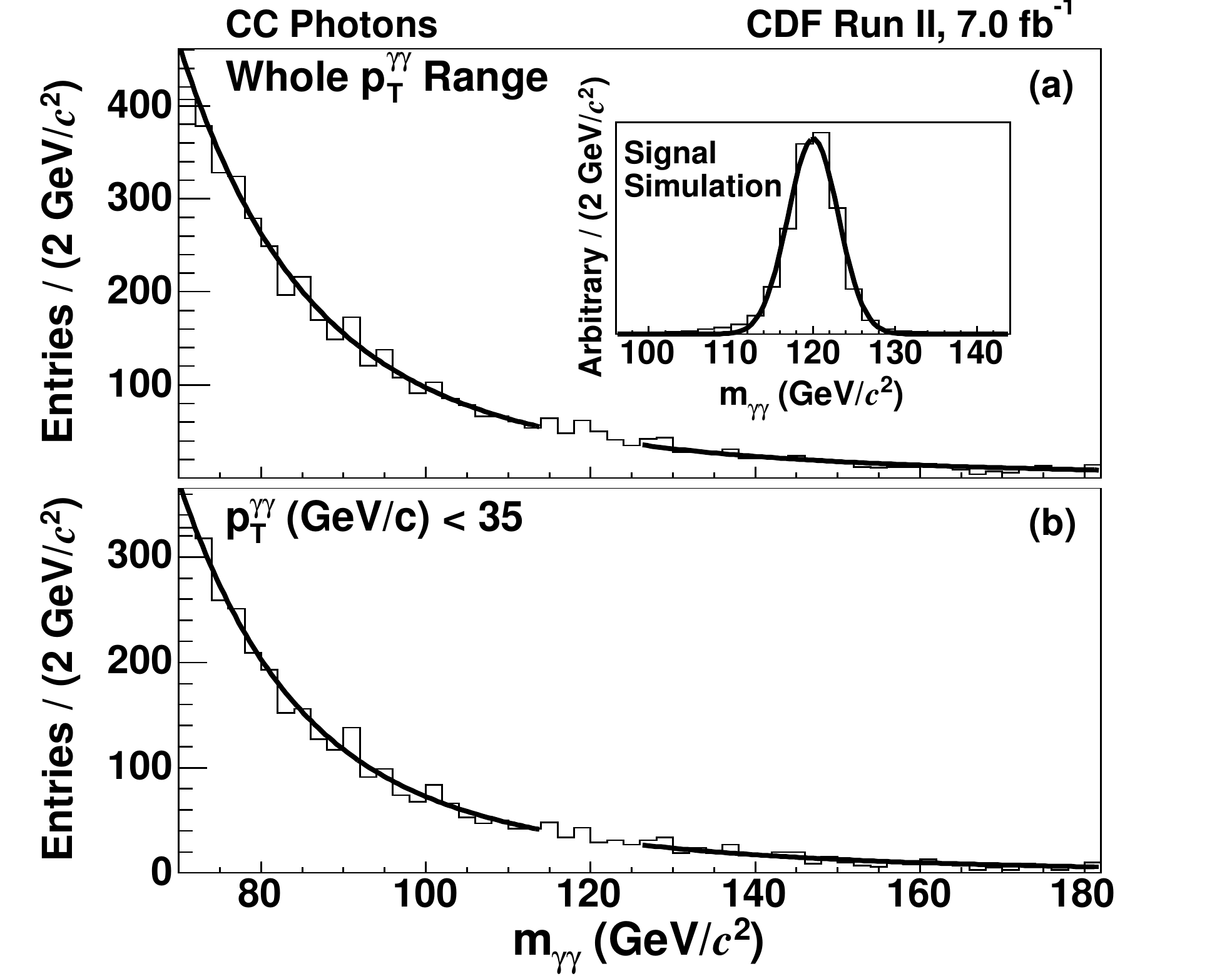}
\includegraphics[scale=0.4]{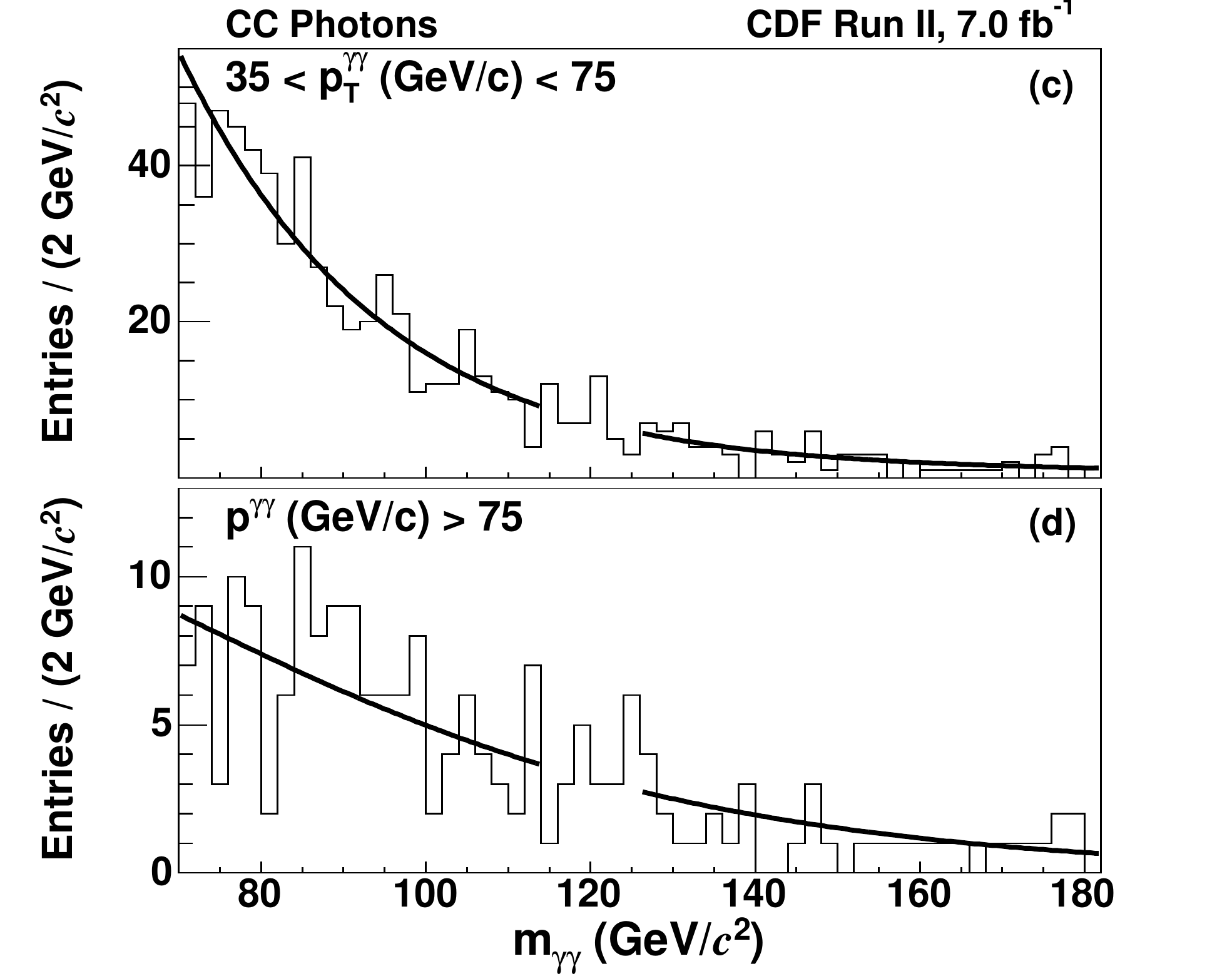}
\end{center}
\caption{As an example, the invariant mass distribution of central-central 
photon pairs is shown for (a) the entire $p_T^{\gamma\gamma}$ region 
used for the SM Higgs boson search
and then divided into
three $p_T^{\gamma\gamma}$ regions (b)--(d) used for the $H_f$ search. 
Each distribution shows a fit to the data for the Higgs boson mass hypothesis of 120~\gevcc. 
The gap in the fit centered at 120~\gevcc\
represents the signal region for this mass point, which was excluded
from the fit.  The expected shape of the signal from simulation is also 
shown in the inset in (a).}  
\label{fig:CDF_fits}
\end{figure}

For the fermiophobic model, the Higgs boson is typically produced in association
with either a $W$ or $Z$ boson or two jets from the VBF process. 
As a result, the fermiophobic Higgs boson has a higher than average 
$p_T^{\gamma\gamma}$ relative to the background processes as it 
is typically recoiling against another object.\footnote{The D0 analysis 
is sensitive to this in both the SM and fermiophobic Higgs boson 
searches by including this variable in the BDT training.}
Therefore, in the fermiophobic scenario, the data are further divided into 
three regions of $p_T^{\gamma\gamma}$, where the highest $p_T^{\gamma\gamma}$
region provides the greatest $H_f$ sensitivity, retaining about 30\% of 
the signal and removing 99.5\% of the background. 
By also including 
the two lower $p_T^{\gamma\gamma}$ regions, a gain in $H_f$ sensitivity 
of about 15\% is obtained compared to using just the higher $p_T^{\gamma\gamma}$
region alone.

At CDF, we use a data-driven background model which
takes advantage of the Higgs boson mass resolution (3~GeV or less) 
and smoothly
falling background in the signal region of the diphoton mass
spectrum. Fits are made to the data excluding a 12 GeV window
centered around each Higgs mass hypothesis. The fit is interpolated
into the signal region to determine the background estimation in that
region and the process is repeated for each subsample 
(CC, CP, C$^{\prime}$C, C$^{\prime}$P and also each
$p_T^{\gamma\gamma}$ region for the fermiophobic Higgs search). 
Fits are performed separately for each 
Higgs boson mass hypothesis
 for 100~$<M_H<$~150~\gevcc\ in 5~\gevcc\ steps.
The statistical uncertainties on the total background
in the signal region, taken from the fit, are considered when setting limits.  
Fits for the CC channel are shown in Figure~\ref{fig:CDF_fits} for the
whole $p_T^{\gamma\gamma}$ region used in the SM search and for
each of the three $p_T^{\gamma\gamma}$ regions used in
the fermiophobic search.

%%%%%%%%%%%%%%%%%%%%%%%%%%%%%%%%%%
\section{Results}

No obvious evidence of a signal is observed in the Tevatron diphoton 
data and BDT discriminants (invariant mass distributions) are used by 
the D0 (CDF) analysis
to set upper limits on the cross section multiplied by the branching ratio 
$\sigma\times B(H\rightarrow\gamma\gamma)$ for both the SM and 
fermiophobic Higgs boson searches at 95\% confidence level (C.L.). 
 Systematic uncertainties on both the predicted
number of signal and background events are considered, in addition
to systematics on the shape of the BDT discriminant for the D0 analysis. 
Correlations between uncertainties are also taken into account.
The D0 analysis uses a modified frequentist approach to 
set upper limits on the Higgs boson production rate and the
CDF analysis uses a Bayesian method. 

\subsection{SM Higgs Boson Search}

\begin{figure}[h!]
\makebox[\textwidth]{
\subfigure{\includegraphics[scale=0.34]{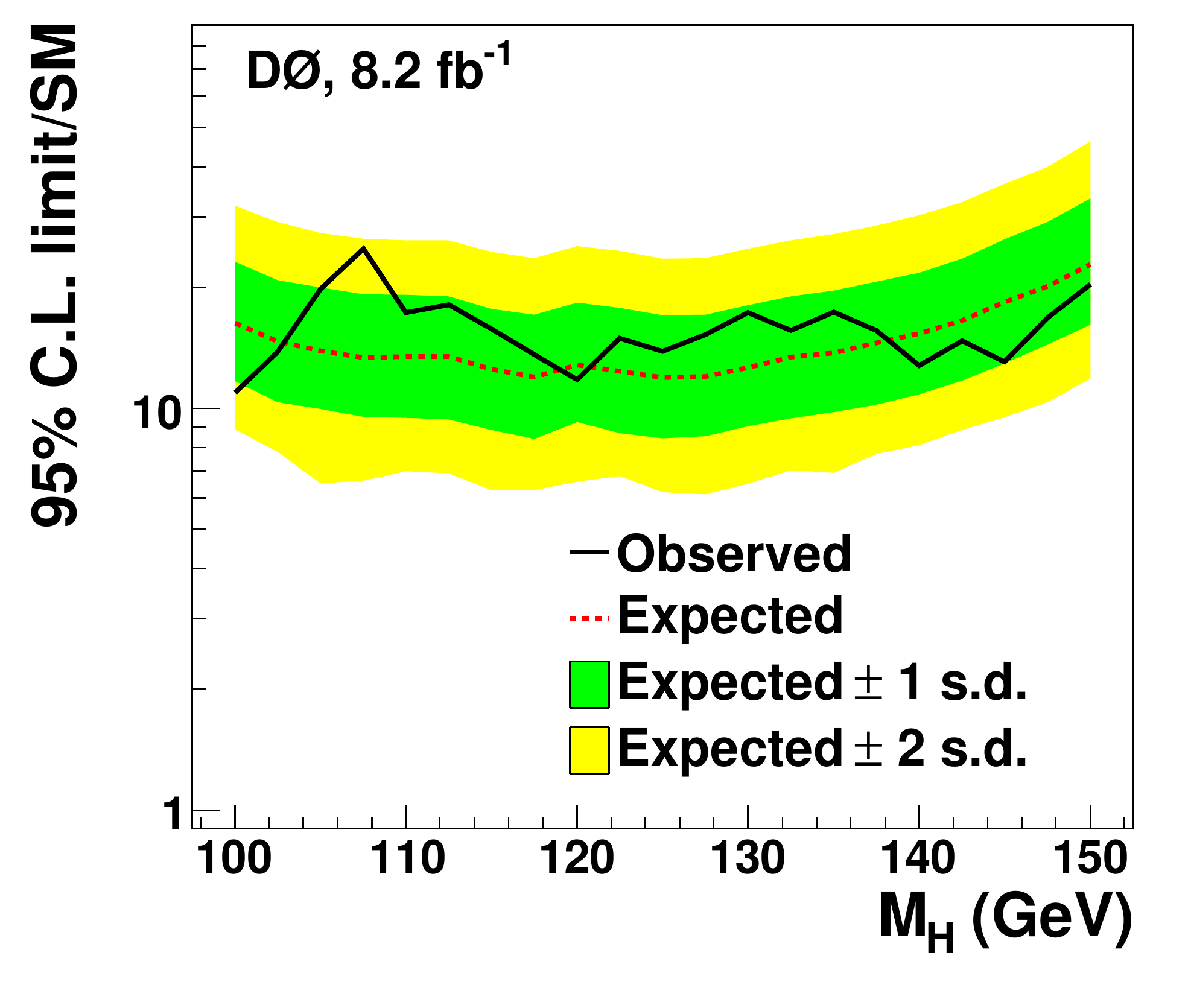}}
\subfigure{\includegraphics[scale=0.39]{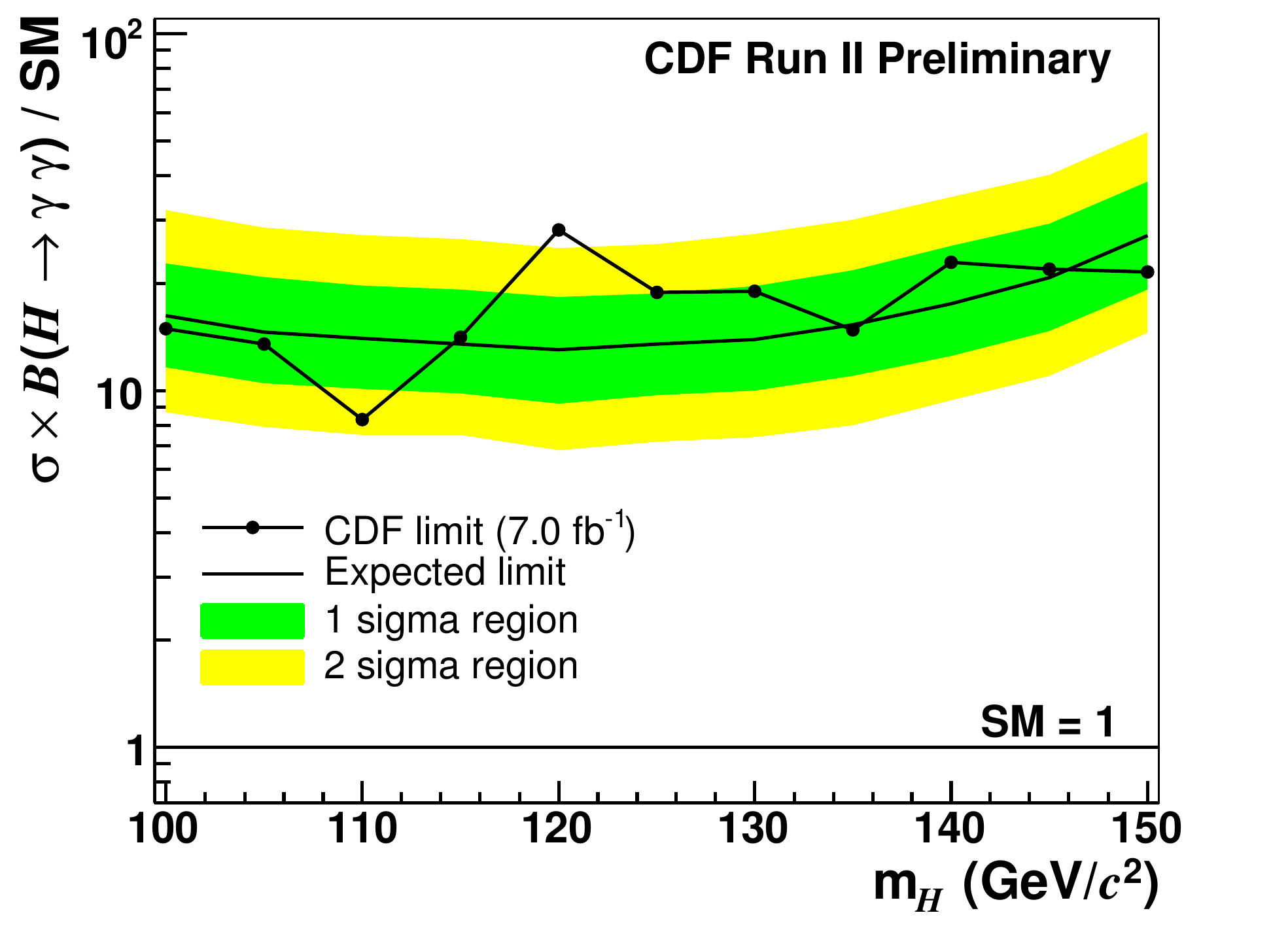}}
}
\caption{Observed and expected 95\% C.L. upper limits on
$\sigma\times B(H\rightarrow\gamma\gamma)$ relative to the SM prediction,
as a function of the SM Higgs boson mass for D0 (left) and CDF (right). The shaded regions represent
the 1$\sigma$ and 2$\sigma$ probability of fluctuations of the observed
limit away from the expected limit based on the distribution of simulated
experimental outcomes under the background-only hypothesis.}
\label{fig:SM_Limits}
\end{figure}

The CDF and D0 observed and expected (median, background-only
hypothesis) limits are shown relative to the SM prediction in 
Figure~\ref{fig:SM_Limits}. The bands indicate the 68\% and 95\%
probability regions where the limits can fluctuate, in the absence of signal.
For the CDF limit at $m_H=$~120~\gevcc, a deviation of greater than two sigma 
from the expectation is observed. 
However, the statistical significance of this discrepancy is reduced below two sigma
after the trial factor associated with performing multiple mass points is 
considered. 

\begin{figure}[h!]
\makebox[\textwidth]{
\subfigure{\includegraphics[scale=0.5]{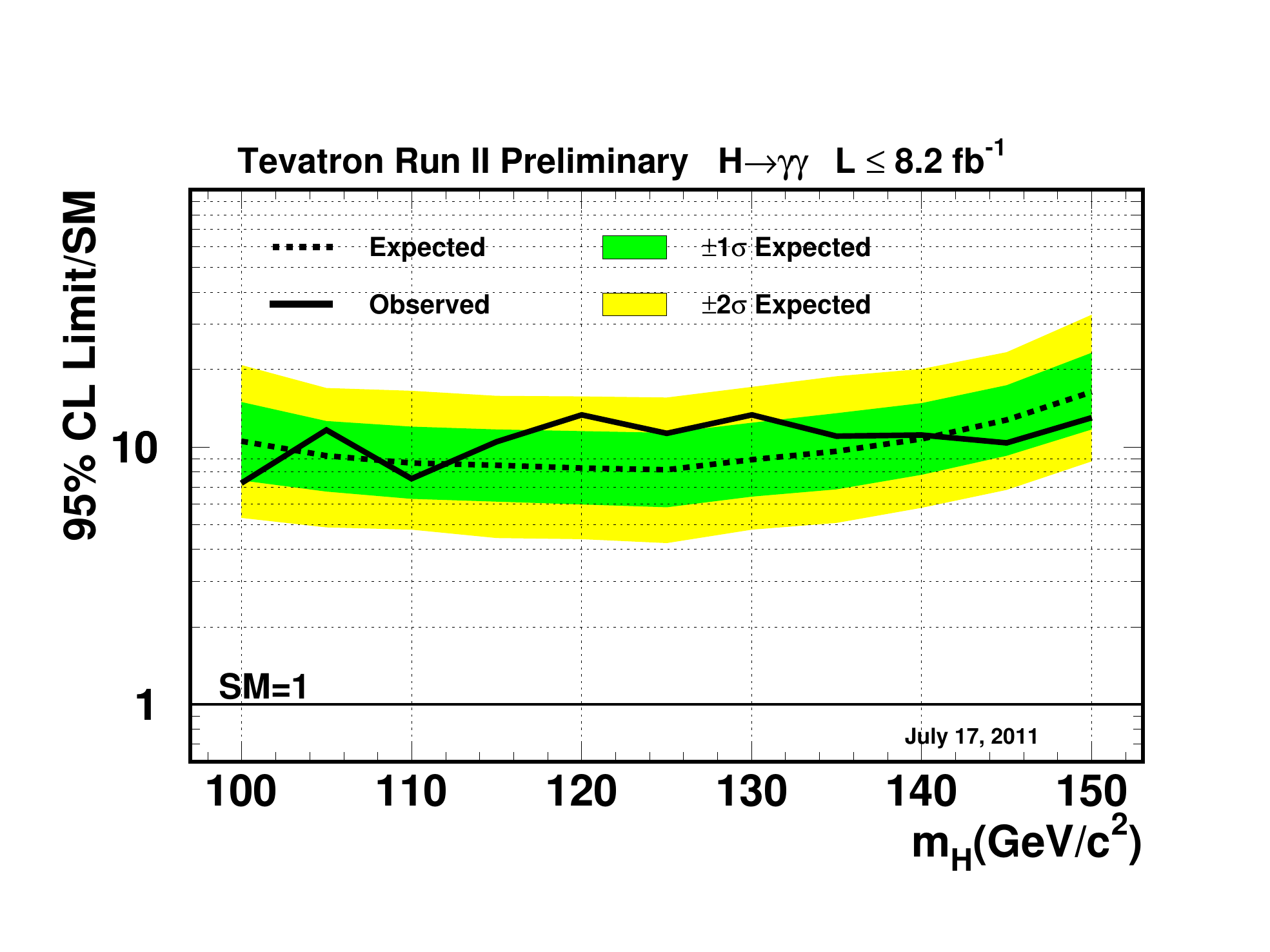}}
\subfigure{\includegraphics[scale=0.5]{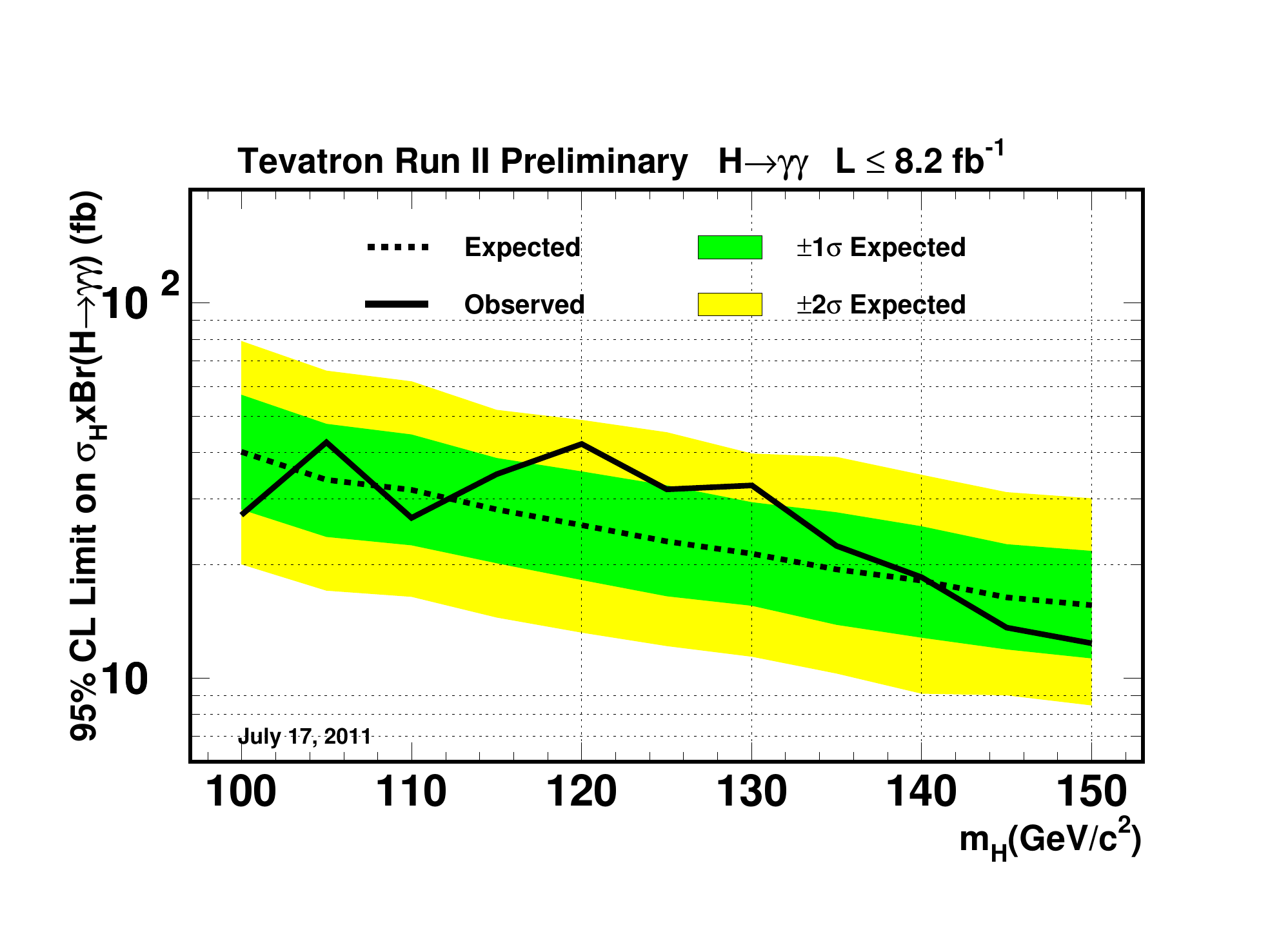}}
}
\caption{At left, limits based on combined D0 and CDF analyses (Fig.~\ref{fig:SM_Limits})
searching for a SM Higgs bosons decaying to two photons.
At right, limits on the inclusive production cross section times the branching ratio to two
photons, with theoretical uncertainties on the cross section removed. The limit results
are calculated using a Bayesian method.}
\label{fig:Tev_Limits}
\end{figure}

Results from the individual searches from CDF and D0 are combined~\cite{Tev_SMHgg_2011}
in order to extract limits on SM Higgs production 
$\sigma\times B(H\rightarrow\gamma\gamma)$
relative to the SM prediction for $100\le M_H \le 150$~\gevcc\ in 5~\gevcc\ steps.
Figure~\ref{fig:Tev_Limits} shows these results obtained from a 
Bayesian method. 
In order to reduce model dependence from the SM predictions, 
limits are also calculated on the inclusive 
cross section times the branching ratio 
$\sigma(p\bar{p}\rightarrow H+X)\times B(H\rightarrow\gamma\gamma)$
with theoretical uncertainties on the total production cross section
removed (Fig.~\ref{fig:Tev_Limits}).

\subsection{Fermiophobic Higgs Boson Search}

\begin{figure}[h!]
\makebox[\textwidth]{
\includegraphics[scale=0.34]{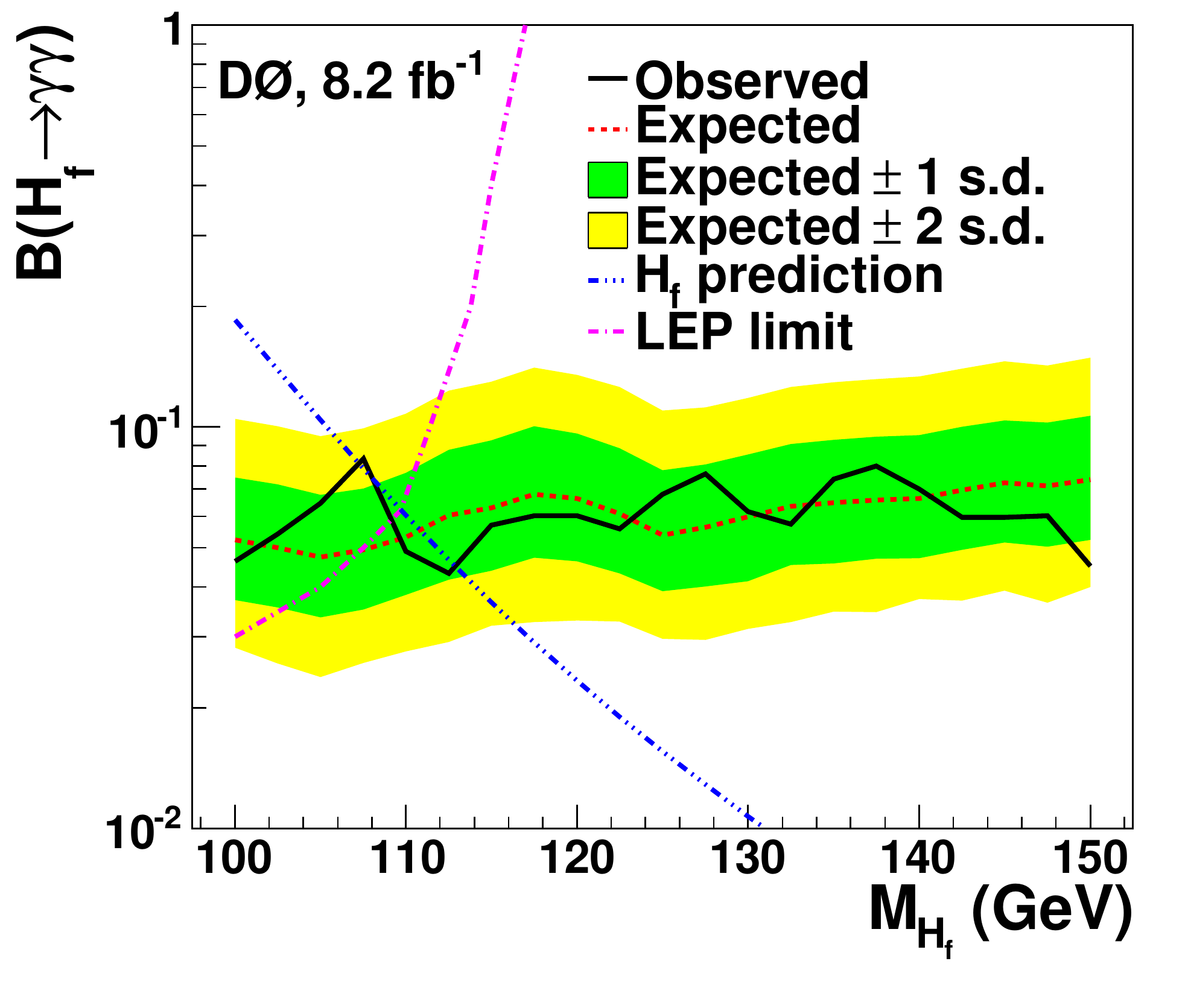}
\includegraphics[scale=0.39]{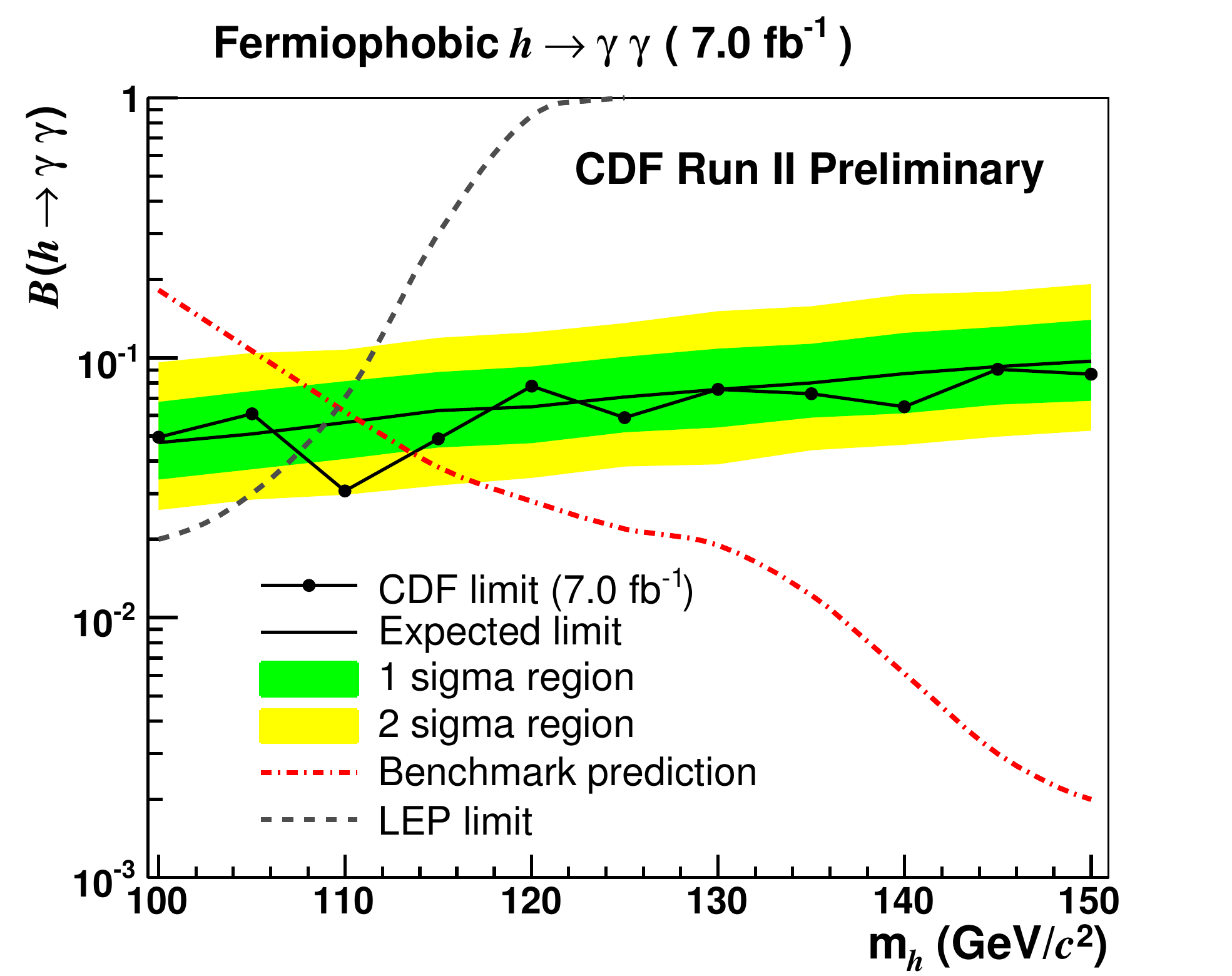}
}
\caption{Observed and expected 95\% C.L. upper limits on fermiophobic $B(H_f\rightarrow\gamma\gamma)$ 
as a function of the Higgs boson mass for D0 (left) and CDF (right).}
\label{fig:BH_Limits}
\end{figure}

In the fermiophobic Higgs boson search, SM cross sections and
uncertainties are assumed with the $gg\rightarrow H$ process excluded,
and the limits on $\sigma\times B(H_f\rightarrow\gamma\gamma)$ 
are converted into limits on $B(H_f\rightarrow\gamma\gamma)$, seen in Fig.~\ref{fig:BH_Limits}. 
Based on an intersection between the observed limit and the model prediction,
fermiophobic Higgs boson masses are excluded below 112.9 (114)~\gevcc\ for
the D0 (CDF) analysis. 
%Each experiment alone, therefore, produces
%more stringent lower limits than that of 109.7~\gevcc\ obtained from
%combined searches at LEP. 

%%%%%%%%%%%%%%%%%%%%%%%%%%%%%%%%%%
\section{Conclusions}

Analyses for the D0 and CDF experiments searched
for a Higgs boson in the diphoton final state using data corresponding to 
8.2 and 7.0~\fb, respectively. 
The results are interpreted in the context of both a SM and fermiophobic Higgs boson
and are found to significantly improve upon the most recent diphoton 
searches from the Tevatron~\cite{hSM_D0, Aaltonen:2009ga}
by more than doubling the amount of data included and by implementing enhanced
search techniques. The D0 analysis uses a neural network to identify photons
with $|\eta|<1.1$ and gains considerable improvement by 
implementing a boosted decision tree technique to better separate 
the Higgs boson signal from the background. The CDF analysis 
now also uses a NN to identify central photons with $|\eta|<1.1$
and additionally gains signal acceptance by including
forward ($1.2<|\eta|<2.8$) and central photon conversions. 

The combined results from these analyses are used to set 95\%
C.L. upper limits on SM Higgs boson production. 
For a Higgs boson mass of $M_H=$~115~\gevcc, 
the observed (expected) limit is a factor of 10.5 (8.5) times the 
SM prediction. These results significantly extend the 
sensitivity of the separate D0 and CDF results and are
the most stringent limits on the SM $H\rightarrow\gamma\gamma$ process 
obtained from Tevatron data. 

In the fermiophobic interpretation, the D0 (CDF) data 
set a lower limit on on the fermiophobic
Higgs boson mass of 112.9 (114)~\gevcc\ at 95\% C.L. 
Each experiment alone, therefore, produces
a more stringent lower limits than that of 109.7~\gevcc\ obtained from
combined searches at LEP.\footnote{See 
Ref.~\cite{Tev_BH_2011} for the Tevatron $H_f$ 
results obtained by combining
the D0 and CDF data (after the DPF conference). This yields a lower limit of $M_{H_f}=$~119 GeV/c$^2$,  
the most stringent limit on the fermiophobic Higgs model.}

\end{document}